\documentclass[fleqn,10pt]{wlscirep}

\usepackage{graphicx} 
\usepackage{amsmath}
\usepackage{bbold}
\DeclareMathAlphabet{\mathbbmsl}{U}{bbm}{m}{sl}

\begin{document}

\title{On the general constraints in single qubit quantum process tomography}
\author[1,*]{Ramesh Bhandari}
\author[2,]{Nicholas A. Peters}
\affil[1]{Laboratory for Physical Sciences, 8050 Greenmead Drive, College Park, Maryland 20740, USA}
\affil[2]{Oak Ridge National Laboratory, One Bethel Valley Road, P.O. Box 2008, MS-6418, Oak Ridge, Tennessee 37831, USA}
\affil[*]{rbhandari@lps.umd.edu}

\affil[ ]{This manuscript has been authored by UT-Battelle, LLC under Contract No. DE-AC05-00OR22725 with the U.S. Department of Energy.  The United States Government retains and the publisher, by accepting the article for publication, acknowledges that the United States Government retains a non-exclusive, paid-up, irrevocable, world-wide license to publish or reproduce the published form of this manuscript, or allow others to do so, for United States Government purposes.  The Department of Energy will provide public access to these results of federally sponsored research in accordance with the DOE Public Access Plan (http://energy.gov/downloads/doe-public-access-plan).}

\date{\today}

\begin{abstract}
We briefly review single-qubit quantum process tomography for trace-preserving and nontrace-preserving processes, and derive  explicit forms of the general constraints for fitting experimental data.  These new forms provide additional insight into the structure of the process matrix. We illustrate their utility with  several examples, including a discussion of qubit leakage error models and the intuition which can be gained from their process matrices.\end{abstract}


\maketitle 

\section*{Introduction}


Despite recent successes in developing new methods such as gate-set tomography (GST)\cite{Merkel, Robin}  to  fully and accurately characterize a given quantum process, as well as simplified methods\cite{heilmann, lu1} to avoid scalability limitations, quantum process tomography (QPT)\cite{1,Poyatos} remains a benchmark standard to which the results of the new evolving methods must be compared.  In this paper, we review single qubit process tomography and present some new findings on the properties of the process matrix in the familiar $\chi$ representation and demonstrate their utility via application to nontrace-preserving processes such as  qubit leakage errors.  In particular, we examine the general form of constraints for numerical fitting of experimental data, and extract simplified forms, which indicate explicit  relationships among the various elements of the  process matrix, one of which is the familiar one, $Tr(\chi)=1$ (in the Pauli basis) for a trace-preserving process.  The other three derived relationships for a trace-preserving process, exclusively involve the off-diagonal elements and thus provide further insight into the structure of the process matrix. Knowledge of these can thus serve as useful tools for an experimentalist interested in measuring quantum gates to determine error models.  We illustrate their utility  with several example process matrices, including some models of leakage errors.


\begin{figure*}
\begin{center}
\includegraphics[width=0.4\textwidth]{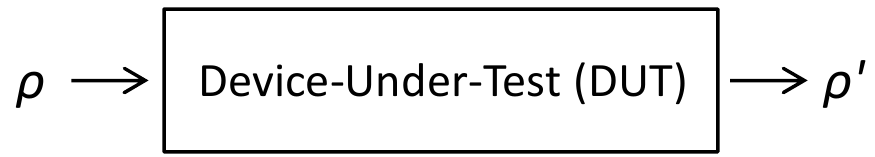}
\caption{The input state $\rho$ changes to $\rho'$ upon traversal through a device-under-test (DUT). \label{fig:one}}
\end{center}
\end{figure*}

Fig.~\ref{fig:one} shows a device under test (DUT) upon which qubits impinge in a quantum state described by the density matrix $\rho$. The output qubits' density matrix is denoted by $\rho'$.  Ordinarily, quantum state tomography produces normalized states; however, the measurement rates contain additional information on the loss to characterize a non-trace-preserving process. To use the loss information, the density matrix of the output state $\rho'$ includes a scaling factor ($\leq 1$) to account for any loss of qubits as they traverse the DUT. 

\section*{Results}

Following\cite{1,NC} , the output state in Figure 1, $\rho'$, 
can then  be written as
\begin{equation}
\rho'=\boldsymbol\epsilon(\rho),  \label{eq:2}
\end{equation}
where $\boldsymbol\epsilon$ is an operator representing the effect of the DUT on the input state. This can be further expanded as \cite{NC}
\begin{equation}
\boldsymbol\epsilon(\rho) =\sum_iE_i\rho E_i^{\dagger}, \label{eq:3}
\end{equation}
where $E_i$'s comprise a set of at most four operators describing the effect of the DUT.  Now these operational elements can be expressed in terms of a \emph{fixed set of basis operators}, $\tilde{E_k}, k=1,2,...4$, i.e., we can write 
\begin{equation}
E_i=\sum_{m=1}^{4}e_{im}\tilde{E}_m, \label{eq:4}
\end{equation}
As a result,
\begin{equation}
\rho'=\sum_{mn}\tilde{E}_m\rho\tilde{E}_n^{\dagger}\chi_{mn}, \label{eq:5}
\end{equation}
where 
$\chi_{mn}=\sum_i e_{im} e_{in}^{*}$. 
Since indices $m$ and $n$ each run from 1 through 4, $\chi_{mn}$ is a 4 x 4 matrix, called the\emph{ process matrix}.
This matrix is  Hermitian. Therefore, it has at most $4^2=16$ independent parameters.  Additionally, it is nonnegative definite, i.e., its eigenvalues are zero or greater.

Now, invoking the fact that for a trace-preserving process,  $Tr(\rho')=1$, one obtains from Eq.~\ref{eq:5} 
\begin{equation}
\sum_{mn}\chi_{mn}\tilde{E}_n^\dagger\tilde{E_m}=\mathbbmsl{I}.  \label{eq:7} 
\end{equation}
These are, in effect, four constraints on the elements, $\chi_{mn}$. These constraints then reduce the number of independent parameters of the $\chi$ matrix from 16 to 12. In general, including nontrace-preserving processes\cite{NC,2} , 
\begin{equation}
P\leq \mathbbmsl{I},\label{eq:8}
\end{equation}
where 
\begin{equation}
P=\sum_i E^\dagger_iE_i=\sum_{mn}\chi_{mn}\tilde{E}_n^\dagger\tilde{E_m}. \label{eq:9}
\end{equation}
Note that the matrix $P$ is nonegative-definite Hermitian.

 In what follows, we choose the Pauli basis, i.e., we set  $\tilde{E}_i=\sigma_i$, where $\sigma_1=\mathbbmsl{I}$, $\sigma_2=\sigma_x$, $\sigma_3=\sigma_y$, and $\sigma_4=\sigma_z$. It can be shown that for this \emph{fixed} set of basis  operators, $Tr(\chi)=Tr(P)/2$, 
which then equals one for a trace-preserving process because in that case, $P=\mathbbmsl{I}$.  Eq.~\ref{eq:8} further implies that the eigenvalues of the $P$ matrix (defined in Eq.~\ref{eq:9}) are each greater than or equal to zero and also less than or equal to one.  For the choice $\tilde{E}_i=\sigma_i$, we find
\begin{equation}
0\leq Tr(\chi) + F \leq 1, \label{eq:10}
\end{equation}
\begin{equation}
0\leq Tr(\chi) - F \leq 1, \label{eq:11}
\end{equation}
where
\begin{equation}
F=2\sqrt{(Im(\chi_{34})+Re(\chi_{12}))^2+(Im(\chi_{24})-Re(\chi_{13}))^2+(Im(\chi_{23})+Re(\chi_{14}))^2}. \label{eq:12}
\end{equation}
 $Tr(\chi) \pm F$ are the two $P$-matrix eigenvalues appearing in the above inequalities, Eqs.~\ref{eq:10} and~\ref{eq:11}, which the $\chi$ matrix must, in general, satisfy (we assume a positive sign for the radical sign in Eq.~\ref{eq:12}).   Adding Eqs.~\ref{eq:10} and~\ref{eq:11} yields $Tr(\chi) \leq 1$, which is normally quoted in literature; however, Eq.~\ref{eq:10} indicates a much tighter constraint, involving both the diagonal elements and the off-diagonal elements. When the process is trace preserving, the equality holds, which then requires that all three terms under the radical sign in Eq.~\ref{eq:12}   be individually equal to zero because $Tr(\chi)=1$.  
In other words, not just $Tr(\chi)=1$,   but the entire set
\begin{equation}
Tr(\chi)=1 \label{eq:13}
\end{equation}
\begin{equation}
Im(\chi_{34})=-Re(\chi_{12}) \label{eq:14}
\end{equation}
\begin{equation}
Im(\chi_{24})=Re(\chi_{13}) \label{eq:15}
\end{equation}
\begin{equation}
 Im(\chi_{23})=-Re(\chi_{14}) \label{eq:16}
\end{equation}
 must  hold in any numerical fit to the experimental data to yield a physical $\chi$ matrix.  To our knowledge, this  explicit form of constraints has not been cited or discussed in the past, although sets of equations of the form, Eq.~\ref{eq:7}, typically have  been employed directly as constraints  in  numerical optimization procedures to obtain a fitted physical (trace-preserving) process matrix from  experimental data (see, e.g.,\cite{Obrien}).   Note that this set of constraints can also be derived directly by solving the linear equations embodied in Eq.~\ref{eq:7}.

From Eq.~\ref{eq:11}, it further follows that $Tr(\chi)\ge F$. Using the fact that both $Tr(\chi$) and $F$ are nonnegative, Eqs.~\ref{eq:10} and~\ref{eq:11} can now be rewritten as 
\begin{equation}
Tr(\chi) + F \leq 1, \label{eq:17}
\end{equation}
\begin{equation}
0\leq Tr(\chi) - F. \label{eq:18}
\end{equation}
These two inequalities serve as general constraints that must be satisfied in a quantum process. 

\section*{Discussion}

Below we give some some examples to corroborate the above results:

\subsection*{Hadamard Gate}

The process matrix for the Hadamard gate is given by
\begin{equation}
\chi_H=\frac{1}{2}
\left[
\begin{matrix}
0&0&0&0\\
0&1&0&1\\
0&0&0&0\\
0&1&0&1
\end{matrix}
\right].\label{eq:19}
\end{equation}
Eq.~\ref{eq:13} is satisfied as $Tr(\chi_H)=1$.  Further there are no complex coefficients, and the first-row elements are all zero, so Eqs.~\ref{eq:14}-\ref{eq:16} are all true and identically zero, as is $F$.   Therefore,  Eqs.~\ref{eq:17} and~\ref{eq:18} are satisfied as well. 

\subsection*{Polarizer at an angle $\theta$}

This is a nontrace-preserving process. The $\chi$ matrix is given by
\begin{equation}
\chi(\theta)=
\left[
\begin{matrix}
1/4&\sin(2\theta)/4&0&\cos(2\theta)/4\\
\sin(2\theta)/4&\sin^2(2\theta)/4&0&\sin(4\theta)/8\\
0&0&0&0\\
\cos(2\theta)/4&\sin(4\theta)/8&0&\cos^2(2\theta)/4
\end{matrix}
\right]. \label{eq:20}
\end{equation}
$Tr(\chi)=1/2$, which is less than 1, as expected.  Additionally, the value of $F$, using Eq.\ref{eq:12}, is also equal to 1/2.  The constraints, Eqs.~\ref{eq:17} and \ref{eq:18}, are satisfied.  Violations occur in Eqs.~\ref{eq:14}-\ref{eq:16}.
\\\\
In addition to $Tr(\chi) <1$ for a nontrace-preserving process,  what specific violations occur in Eqs.~\ref{eq:14}-\ref{eq:16}  can also be an indication of the type of nontrace-preserving process. We illustrate this with respect to a leakage error model for quantum computing. 

\subsection*{Leakage Error Model}

Qubit leakage is  of two types: 1) coherent leakage, where the qubit represented by a two-level subsystem of a multi-level system like the trapped ion, leaks out of its Hilbert space and then transitions back to it; 2) loss, where the qubit permanently transitions out of its Hilbert space, i.e., never returns to it and is thus considered lost. In this paper, we focus on the latter, where, for example, the qubit  in the first excited state ($|1\rangle$) of the multi-level system, may be further excited outside of the qubit's computational Hilbert space, and never returns to it (or returns to it after a very long time, so for practical purposes it is considered lost). The process is therefore nontrace preserving. 
Following\cite{Wall} , 
\begin{equation}
\rho'=\mathcal{E}(\rho)= (1-p)\rho+\frac{p}{4}(\mathbbmsl{I}+\sigma_z)\rho(\mathbbmsl{I}+\sigma_z), \label{eq:21}
\end{equation}
where $\mathcal{E}$ represents the error operation,  $\rho$ is the input state, $\rho'$ is the output state and $p$ is the leakage error probability. It  follows from above  that
\begin{equation}
Tr(\rho') = (1-\frac{p}{2})+\frac{p}{2}Tr(\rho\sigma_z), \label{eq:22}
\end{equation}
indicating that the qubit is lost with a probability $p$ when it is in the excited state and remains stable when it is in the ground state ($|0\rangle$).  Using Eq.~\ref{eq:5} and Eq.~\ref{eq:21}, the process matrix  is 
\begin{equation}
\chi_\mathcal{E_Z}=
\left[
\begin{matrix}
1-\frac{3p}{4}&0&0&\frac{p}{4}\\
0&0&0&0\\
0&0&0&0\\
\frac{p}{4}&0&0&\frac{p}{4}
\end{matrix}
\right].\label{eq:23}
\end{equation}
$Tr(\chi_\mathcal{E_Z}) =1- \frac{p}{2}\leq 1$ for $p \geq 0$. In this case, when $p>0$, this is no longer a trace-preserving process, so Eq.~\ref{eq:16} is violated in proportion to the leakage probability $p$. In fact, all the nonzero, non-identity elements deviate from the corresponding elements of the ideal identity gate by an amount identical in magnitude ($p/4$), which is  proportional to the leakage probability $p$.

Consider now the case where in Eq.~\ref{eq:21}, the Pauli operator, $\sigma_z$ is replaced by $\sigma_x$. This is a nontrace-preserving process with $Tr(\mathcal{E}(\rho))$ given by Eq.~\ref{eq:22}, but with $\sigma_z$ replaced with $\sigma_x$. This corresponds to a noisy environment where the state $|+\rangle \equiv (|0\rangle +|1\rangle )/\sqrt{2}$ stays stable, and the state $|-\rangle\equiv (|0\rangle - |1\rangle )/\sqrt{2}$ leaks out with probability $p$.  On the other hand, $|0\rangle$ and $|1\rangle$, which comprise the $|+\rangle$ and the $|-\rangle$ states,  leak out with the same probability, $1-p/2$. The corresponding process matrix is given by
\begin{equation}
\chi_\mathcal{E_X}=
\left[
\begin{matrix}
1-\frac{3p}{4}&\frac{p}{4}&0&0\\
\frac{p}{4}&\frac{p}{4}&0&0\\
0&0&0&0\\
0&0&0&0 
\end{matrix}
\right]. \label{eq:24}
\end{equation}
Here  the violation, indicative of a nontrace-preserving process, occurs in Eq.~\ref{eq:14}, instead of Eq.~\ref{eq:16}, signifying a different nontrace-preserving process, even though $1-Tr(\chi)$  remains unchanged.  The positioning of the nonzero elements, except the first diagonal element here, has  shifted within the $\chi$ matrix, suggestive of the change in the nature of the nontrace-preserving process.  
This manner of shift is predictable if one is specifically working with a general leakage error model in which $\sigma_z$ in Eq.~\ref{eq:21} is replaced with $\vec{\sigma}.\vec{n}$.   

We further extend the model of Eq.~\ref{eq:21} to qubits, where the ground state ($|0\rangle)$ may also leak out, although with  a  low  probability compared to the excited state ($|1\rangle$) as, for example, in superconducting phase qubits\cite{Martinis} .   The leakage process here can be represented by  the following equation:
\begin{equation}
\mathcal{E}(\rho)= (1-p)\rho+\frac{p}{4}(\mathbbmsl{I}+\sigma_z)\rho(\mathbbmsl{I}+\sigma_z) - \frac{\epsilon}{2} (\rho \sigma_z+ \sigma_z\rho), \label{eq:25}
\end{equation}
where an extra term has been added to Eq.~\ref{eq:21} to account for the leakage of the ground state as seen below:
\begin{equation}
Tr(\mathcal{E}(\rho)) = (1-\frac{p}{2})+\frac{p-2\epsilon}{2}Tr(\rho\sigma_z). \label{eq:26}
\end{equation}
The ground state leakage probability, from Eq.~\ref{eq:26}, is $\epsilon$.  While the excited state leakage probability is $p-\epsilon$.  The process matrix is the same as the one given in Eq.~\ref{eq:23}, except that the nonzero off-diagonal elements are now changed to $(p-2\epsilon)/4$, an indication of the change of the nature of the nontrace-preserving process, namely, the presence of leakage from the ground state as well.  We also note here that the left hand side of Eq. \ref{eq:17},  $Tr(\chi)+F$ evaluates to $1-\epsilon$ for this model in contrast to the value of 1 obtained for Eqs.~\ref{eq:23} and \ref{eq:24}, which can be another distinguishing feature.

Thus, we see that simplification of the constraints, Eq.~\ref{eq:9}  into  the set,  Eqs.~\ref{eq:13}-\ref{eq:16}, can provide  insight into the structure of the trace-preserving process matrix;  the three newly derived explicit forms, Eqs.~\ref{eq:14}-\ref{eq:16}, express clear relationships among the off-diagonal elements; we have not seen these relationships mentioned or discussed in the literature before. Violations of these constraints is an indication of a nontrace-preserving process, and the nature of the violations, as we have illustrated above, can help discriminate one type of a nontrace-preserving process from another.  Furthermore, it must be emphasized that for a quantum process known to be nontrace-preserving like the polarizer (where $Tr(\chi) =1/2$, ideally), or for a process suspected to be not strictly trace-preserving like a  quantum gate with leakage errors, or simply for a DUT whose behavior is not known a priori (a true black box), the general constraints, Eqs.~\ref{eq:17}~and~\ref{eq:18},  must be invoked in the fitting of data.

In summary, we have revisited the theoretical aspects of single qubit quantum process tomography to determine the  behavior of  a quantum device. More specifically, we have reexamined the well-known constraints for the process matrix (in the $\chi$ representation), and recast them into more insightful forms.  In the case of a trace-preserving process, specific relationships among the various elements of the process matrix emerge that then shed light on its basic  generic structure. Knowledge of these new constraint relationships   permit an enhanced understanding of the interpretation and analysis of the experimental data.   We have illustrated their validity and utility with several examples, with specific attention to leakage errors, which are of significant importance in quantum computing.  

\section*{Methods}

We tested the efficacy of constraints, Eqs.~\ref{eq:17} and~\ref{eq:18}, in fitting data by adding noise to the above ideal $\chi$ matrices for the Hadamard gate,  the polarizer, and the leakage error models considered in this paper. We simulated Gaussian Hermitian complex noise  using the MATLAB R2015b function randn which returns a number from a normal distribution with zero mean and a standard deviation equal to 1.  This noise is then scaled by a variable scaler ranging from $10^{-4}$ to $10^{-1}$ and added to the process matrix, after which the process matrix is optimized; one fixed value of the scaler is used at one time.  We use toolboxes YALMIP Version 19-Sep-2015~\cite{YALMIP} with SeDuMi 1.32~\cite{SEDUMI} for optimization within Matlab.

In the numerical simulations,  we  frequently observed the noisy $\chi$ matrices  to have negative eigenvalues, eigenvalues exceeding unity, and/or trace exceeding unity.  Imposing the requirements of nonnegative definiteness, Hermiticity and the constraints, Eqs.~\ref{eq:17} and~\ref{eq:18} to fit these noisy $\chi$ matrices always restored  physicality; the eigenvalues were then  nonnegative and less than or equal to 1.  Improperly constraining the system, e.g., imposing only $Tr(\chi)\leq 1$,  without Eqs.~\ref{eq:17} and~\ref{eq:18},  led to unphysical output states computed from $\chi$, even though the requirements of nonnegative definiteness and Hermiticity for the $\chi$ matrix were still in place.  Further it is worth noting that in many examples examined, the fidelity between the target process matrix and each of the two types of optimizations is similar, especially when it is high, and in this case does not aid one in detecting optimization errors.  

Next we give two specific examples showing an initial noisy process matrix and the results after applying  the complete constraints.  First we consider the Hadamard gate as given by Eq.~\ref{eq:19}.  After adding noise scaled by $10^{-3}$, we obtain, as an example, the following:

\begin{equation}
\chi_{H initial}=
\left[
\begin{matrix}
   -0.0009 + 0.0000i  &-0.0005 - 0.0007i  &-0.0012 - 0.0002i  &-0.0003 + 0.0015i\\
  -0.0005 + 0.0007i   &0.4998 + 0.0000i  &-0.0011 + 0.0016i   &0.5012 - 0.0002i\\
  -0.0012 + 0.0002i  &-0.0011 - 0.0016i  &-0.0015 + 0.0000i  &-0.0004 - 0.0002i\\
  -0.0003 - 0.0015i  &0.5012 + 0.0002i  &-0.0004 + 0.0002i   &0.5003 + 0.0000i
\end{matrix}
\right]. \label{eq:m1}
\end{equation}
This initial matrix has one eigenvalue greater than one and two negative eigenvalues and is therefore unphysical. We also note that the set of Eqs. \ref{eq:13}-\ref{eq:16} is violated here.   Here and in the following examples, we show rounded results, while full precision is used to compute reported derived quantities.  

Under the assumption of a trace-preserving process, we perform numerical fitting using Eqs.  \ref{eq:13}-\ref{eq:16}, as constraints. The result is
\begin{equation}
\chi_{H_{TP}}=
\left[
\begin{matrix}
   0.0000 + 0.0000i   &0.0000 + 0.0004i  &-0.0000 - 0.0000i  &-0.0000 + 0.0004i\\
   0.0000 - 0.0004i   &0.4997 + 0.0000i  &-0.0008 + 0.0000i   &0.5000 - 0.0000i\\
  -0.0000 + 0.0000i  &-0.0008 - 0.0000i   &0.0000 + 0.0000i  &-0.0008 - 0.0000i\\
  -0.0000 - 0.0004i   &0.5000 + 0.0000i  &-0.0008 + 0.0000i   &0.5003 + 0.0000i
\end{matrix}
\right]. \label{eq:m2}
\end{equation}
Eqs. \ref{eq:13}-\ref{eq:16} are now  satisfied. If, on the other hand, the quantum process is suspected to be not strictly trace-preserving (due to the possibility of leakage errors), one must replace the constraints, Eqs. \ref{eq:13}-\ref{eq:16}, with the the general constraints, Eqs.\ref{eq:17} and \ref{eq:18}. The result, after fitting with these constraints, is

\begin{equation}
\chi_{H_{NTP}}=
\left[
\begin{matrix}
   0.0000 + 0.0000i  &-0.0000 + 0.0004i  &-0.0000 - 0.0000i  &-0.0000 + 0.0004i\\
  -0.0000 - 0.0004i   &0.4997 + 0.0000i  &-0.0008 + 0.0000i   &0.5000 - 0.0000i\\
  -0.0000 + 0.0000i  &-0.0008 - 0.0000i   &0.0000 + 0.0000i  &-0.0008 - 0.0000i\\
  -0.0000 - 0.0004i  &0.5000 + 0.0000i  &-0.0008 + 0.0000i   &0.5002 + 0.0000i
\end{matrix}
\right]. \label{eq:m3}
\end{equation}
Eq., \ref{eq:m2} and \ref{eq:m3} are very similar, however, the latter's trace is 0.9999, so it is not trace preserving, but it is a valid physical process.  

As a second example, we consider the leakage error model described by  Eq.~\ref{eq:25} (a nontrace-preserving process) with $p = 10^{-2}$, $\epsilon = 3*10^{-3}$, and the Gaussian noise scaler equal to $10^{-3}$.  An instance of the noisy process matrix is
\begin{equation}
\chi_{initial}=
\left[
\begin{matrix}
    0.9921 + 0.0000i   &0.0012 - 0.0012i  &-0.0032 - 0.0011i   &0.0013 + 0.0006i\\
   0.0012 + 0.0012i   &0.0004 + 0.0000i   &0.0001 - 0.0002i  &-0.0016 + 0.0008i\\
  -0.0032 + 0.0011i   &0.0001 + 0.0002i  &-0.0022 + 0.0000i   &0.0013 + 0.0006i\\
   0.0013 - 0.0006i  &-0.0016 - 0.0008i   &0.0013 - 0.0006i   &0.0042 + 0.0000i
\end{matrix}
\right]. \label{eq:m4}
\end{equation}
It has two negative eigenvalues, and is therefore unphysical.   In addition, Eq.~\ref{eq:17} is violated as the left-hand side evaluates to a value of 1.0034.  After optimization with constraints, Eqs.\ref{eq:17} and \ref{eq:18}, the process matrix is 
\begin{equation}
\chi_{NTP}=
\left[
\begin{matrix}
    0.9911 + 0.0000i   &0.0009 - 0.0012i  &-0.0023 - 0.0011i   &0.0009 + 0.0006i\\
   0.0009 + 0.0012i   &0.0004 + 0.0000i  &-0.0002 - 0.0000i  &-0.0012 - 0.0002i\\
  -0.0023 + 0.0011i  &-0.0002 + 0.0000i   &0.0001 + 0.0000i   &0.0006 + 0.0000i\\
   0.0009 - 0.0006i  &-0.0012 + 0.0002i   &0.0006 - 0.0000i   &0.0034 + 0.0000i
\end{matrix}
\right]. \label{eq:m5}
\end{equation}
The optimized result is nonnegative definite and satisfies the required constraints, Eqs.  \ref{eq:17} and \ref{eq:18}. 

\section*{Acknowledgements}

One of us (NAP) acknowledges research sponsored by the Laboratory Directed Research and Development Program of Oak Ridge National Laboratory, managed by UT-Battelle, LLC, for the U. S. Department of Energy.   RB acknowledges useful communications with Joel Wallman, Joseph Emerson, Andrzej Veitia, and Robin Blume-Kahout.

\section*{Author contributions statement}
RB and NAP contributed to the conception, implementation, and analysis of these results. All authors have reviewed the manuscript.

\section*{Additional information}
\textbf{Competing financial interests} The authors declare no competing financial interests.


\end{document}